\documentclass[galley,grl]{agu2001}

\usepackage{graphicx}
\authorrunninghead{DAVIDSEN, GRASSBERGER AND PACZUSKI}

\titlerunninghead{Earthquake recurrence}

\authoraddr{J. Davidsen, British Antarctic Survey, High Cross, Madingley
Road, Cambridge CB3 0ET, UK. (j.davidsen@bas.ac.uk)}
\begin{document}

%
%

\title{Earthquake recurrence as a record breaking process}

%
%


\author{J\"orn Davidsen}
\affil{Max-Planck-Institut f\"ur Physik Komplexer Systeme,
Dresden, Germany}
\affil{Perimeter Institute for Theoretical Physics, Waterloo, Canada, N2L
2Y5}

\author{Peter Grassberger}
\affil{John-von-Neumann Institute for Computing, FZ J\"ulich, 52425 J\"ulich, Germany}
\affil{Perimeter Institute for Theoretical Physics, Waterloo, Canada, N2L
2Y5}

\author{Maya Paczuski}
\affil{Perimeter Institute for Theoretical Physics, Waterloo, Canada, N2L
2Y5}
\affil{Complexity Science Group, Department of Physics and Astronomy, University
of Calgary, Calgary, Alberta, Canada T2N 1N4}

%
%

\begin{abstract}

Extending the central concept of recurrence times for a point
process to recurrent events in space-time allows us to
characterize seismicity as a record breaking process using only
spatiotemporal relations among events. Linking record breaking
events with edges between nodes in a graph generates a complex
dynamical network isolated from any length, time or magnitude
scales set by the observer. For Southern California, the network
of recurrences reveals new statistical features of seismicity with
robust scaling laws. The rupture length and its scaling with
magnitude emerges as a generic measure for distance between
recurrent events. Further, the relative separations for subsequent
records in space (or time) form a hierarchy with unexpected
scaling properties.

\end{abstract}

%
%

\begin{article}

\section{Introduction}

Fault systems as the San Andreas fault in California or the Sunda
megathrust (the great tectonic boundary along which the Australian
and Indian plates begin their descent beneath Southeast Asia) are
prime examples of self-organizing systems in nature
\citep{rundle02}. Such systems are characterized by interacting
elements, each of which stays quiescent in spite of  increasing
stress acting on it until the stress reaches a trigger threshold
leading to a rapid discharge or "firing". Since the internal state
variables evolve in time in response to external driving sources
and inputs from other elements, the firing of an element may in
turn trigger a discharge of other elements. In the context of
fault systems, this corresponds to earthquakes, or the deformation
and sudden rupture of parts of the earth's crust driven by
convective motion in the mantle.

Fault systems --- and driven threshold systems in general ---
exhibit dynamics that is strongly correlated in space and time
over many scales. Their complex spatiotemporal dynamics manifests
itself in a number of generic, empirical features of earthquake
occurrence including clustering, fault traces and epicenter
locations with fractal statistics, as well as  scaling laws like
the Omori and Gutenberg-Richter (GR) laws (see e.g. Refs.
\citep{turcotte,rundle03} for a review), giving rise to a
worldwide debate about their explanation. Resolving this dispute
could conceivably require measuring the internal state variables
--- the stress and strain everywhere within the earth along
active faults --- and their exact dynamics. This is (currently)
impossible. Yet, the associated earthquake patterns are readily
observable making a statistical approach based on the concept of
spatiotemporal point processes feasible, where the description of
each earthquake is reduced to its size or magnitude, its epicenter
and its time of occurrence. Describing the patterns of seismicity
may shed light on the fundamental physics since these patterns are
emergent processes of the underlying many-body nonlinear system.

Recently, such an approach has brought to light new properties of
the clustering of seismicity in space and time
\citep{bak02,corral03,corral04,davidsen04,davidsen05m,baiesi_pac},
which can potentially be exploited for earthquake prediction
\citep{goltz01,tiampo02,baiesi}. One aim has been to evaluate
distances between subsequent events, including temporal and
spatial measures. The observed spatiotemporal clustering of
seismicity suggests that subsequent events are to a certain extent
causally related. It further suggests that the usual
mainshock/aftershock scenario --- where each event has at most one
correlated predecessor --- is too simplistic and that the causal
structure of seismicity could extend beyond immediately subsequent
events, especially since the determination of the sequence is
largely arbitrary depending on the size of the region considered
and the completeness of the record of events.

In this work we quantify the spatiotemporal clustering of
seismicity in terms of a sparse, directed network, where each
earthquake is a node in the graph and links connect events with
their recurrences. This general network picture allows us to
characterize clustering by using only the spatiotemporal structure
of seismicity, without any additional assumptions.

\section{Method}

The key advance we propose is to generalize the notion of a
subsequent event to a record breaking event, one which is closer
in space than all previous ones, up to that time. Consider a pair
of events, $A$ and $B$, occurring at times $t_A < t_B$. Earthquake
$B$ is a recurrence of $A$ -- or record with respect to $A$ -- if
no intervening earthquake happens in the spatial disc centered on
$A$ with radius $\overline{AB}$ during the time interval
$[t_A,t_B]$.  Each recurrence is characterized by the distance
$l=\overline{AB}$ and the time interval $T=t_B-t_A$ between the
two events. Since the spatial window is centered on the first
event, any later recurrence to it is closer in space than all
previous ones, and for that reason constitutes another record
breaking event.
\footnote{Notice the difference to the definition of an
$\epsilon$-recurrence, where any event $B$ is considered a
recurrence of $A$ if it occurs at a spatial distance less than
some fixed threshold $\epsilon$ \citep{eckmann87}. In our
definition we do not impose any threshold but allow the sequence
of events themselves to determine which events are recurrences to
other ones.}
This gives rise to a hierarchical cascade of recurrences, where
each recurrence is, by construction, a record. Note that each
earthquake has its own sequence of records or recurrences that
follow it in time.

Our definition of recurrent events is based solely on
spatiotemporal relations between events and minimizes the
influence of the observer by avoiding the use of any space, time,
or magnitude scales other than those explicitly associated with
the earthquake catalog (i.e. its magnitude, spatial, and temporal
ranges). Even the influence of the later scales is rather small
since, for example, an increase in the spatial-temporal coverage
of the catalog does not generally turn a record-breaking event in
a non-record breaking event, thus, conserving the property of a
record. Our definition further allows us to discuss spatial and
temporal clustering, without introducing any artificial scales, or
making any arbitrary assumptions about the form of seismic
correlations. Also, as time goes on, one wants to be more strict
in declaring $B$ a recurrence of $A$, or related to $A$ in a
meaningful way, which is precisely what our definition achieves.

To construct a network we represent each earthquake as a node, and
each recurrence by a link between pairs of nodes, directed
according to the time ordering of the earthquakes.  Distinct
events can have different numbers of in-going and out-going links,
which designate their relations to the other events. The out-going
links from any node define the structure of recurrences in its
neighborhood and characterize the spatiotemporal dynamics of
seismicity, or its clustering with respect to that event.  The
overall structure of the network describes the clustering of
seismic activity in the region that is analyzed.

To test the suitability and robustness of our method to
characterize seismicity, we study a ``relocated" earthquake
catalog from Southern California
\footnote{http://www.data.scec.org/ftp/catalogs/SHLK/}
which has improved relative location accuracy within groups of
similar events, the relative location errors being less than 100m
\citep{shearer03}. The catalog is assumed to be homogeneous from
January 1984 to December 2002 and complete for events larger than
magnitude $m_c=2.5$ \citep{wiemer00}. Restricting ourselves to
epicenters located within the rectangle $(120.5^\circ W,
115.0^\circ W)\times(32.5^\circ N, 36.0^\circ N)$ and to
magnitudes $m \geq m_c$ gives $N = 22217$ events. In order to test
for robustness and the dependence on magnitude, we analyze this
sub-catalog and subsets of it that are obtained by selecting
higher threshold magnitudes, namely $m=3.0, 3.5, 4.0$ giving
$N=5857, 1770, 577$ events, or a shorter period from January 1984
to December 1987 giving $N=4744$ events for $m=m_c$.

\section{Results \& Discussion}

\begin{figure}
\noindent\includegraphics*[width=20pc]{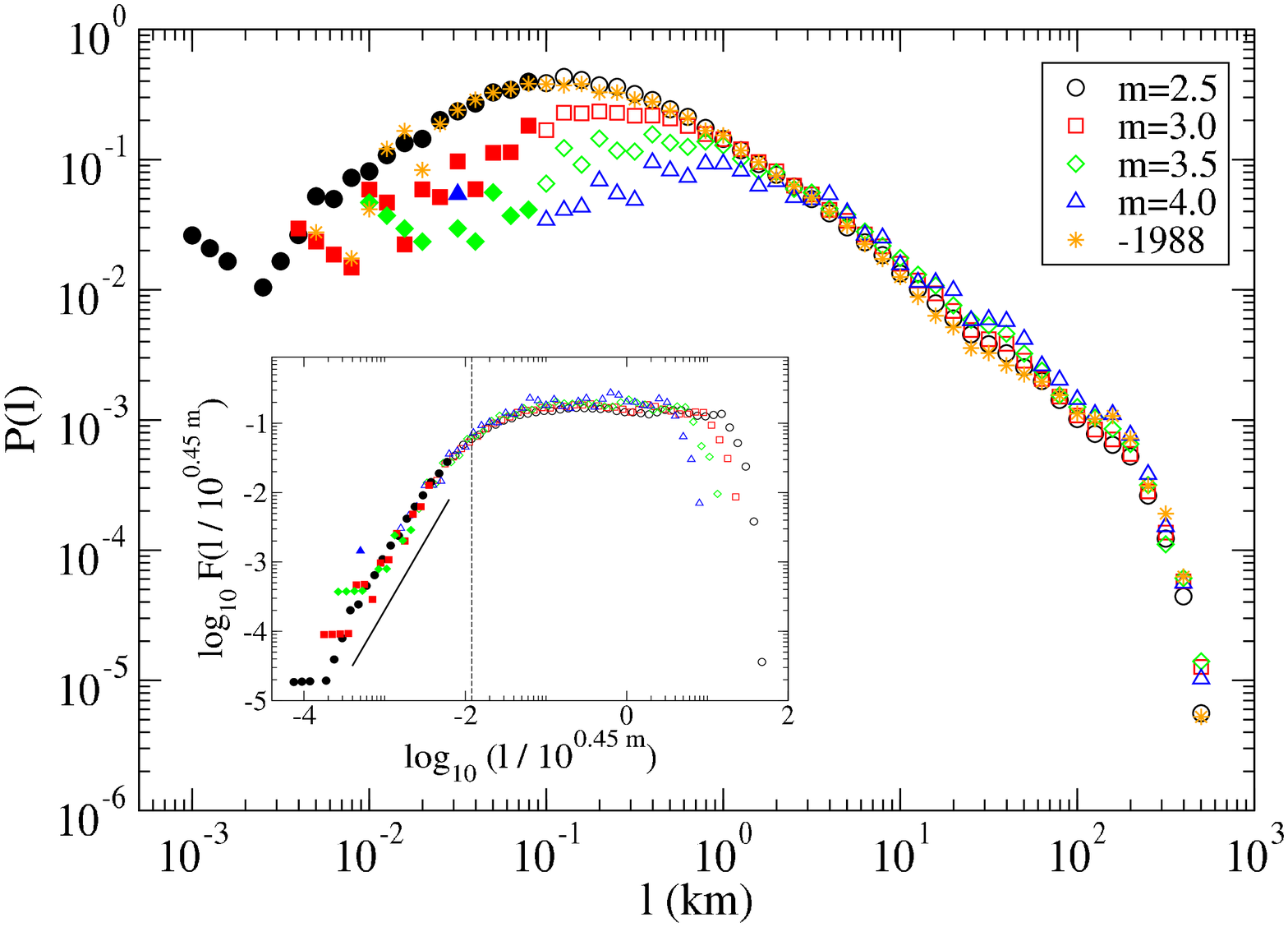}
\caption{\label{fig:l} Distribution of distances $l$ of recurrent
events for sets with different magnitude thresholds $m$. The
distribution for $m=2.5$ up to 1988 is also shown. Filled symbols
correspond to distances below 100~m and are unreliable due to
location errors. The inset shows a data collapse, obtained by
rescaling distances and distributions according to
Eq.~\ref{equ:p(l)}. The full straight line has slope 2.05; the
vertical dashed line indicates the pre-factor $L_0$ in the scaling
law for the characteristic distance, $l^* = L_0 \times
10^{0.45m}$. }
\end{figure}

Fig.~\ref{fig:l} shows the probability distribution function
$P_m(l)$ of distances, $l$, of recurrent events for different
thresholds $m$. The typical or characteristic distance, $l^*(m)$,
where the distribution peaks,  increases with magnitude. For
sufficiently large $l$, all distributions show a power law decay
with an exponent $\approx 1.05$ up to a cutoff. This cutoff is the
size of the region of Southern California that we consider.

With a suitable scaling ansatz, the different curves in
Fig.~\ref{fig:l} fall onto a universal curve, except at the
cutoff, which is a man-made scale imposed on the geological
system. The inset in Fig.~\ref{fig:l} shows results of a data
collapse using
\begin{equation}
\label{equ:p(l)}
 P_m(l) \sim
   l^{-1.05}F(l/10^{0.45m}) \; .
\end{equation}
The scaling function $F$ has two regimes, a power-law increase
with exponent $\approx 2.05$ for small arguments and a constant
regime at large arguments. The transition point between the two
regimes can be estimated by extrapolating them and selecting the
intersection point, giving $L_0 = 0.012$km. For the characteristic
distance that appears in $F$ we thus find $l^* \approx L_0 \times
10^{0.45m}$. This is close to the estimated behavior of the
rupture length $L_R \approx 0.02 \times 10^{m/2}$ km given by
\cite{kagan02} and remarkably close to $L_R = \sqrt{A_R} \approx
0.018 \times 10^{0.46\;m}$ km given by \cite{wells94}, where $A_R$
is the rupture area.

The agreement between our result and that of \cite{wells94}
suggests that the characteristic length scale of distances of
recurrent events is the rupture length, defined in terms of the
rupture area $l^*=L_R\equiv\sqrt{A_R}$. This is substantially
supported by the remarkable fact that, for fixed $m$, $P_m(l)$ and
thus $l^*$ does not significantly vary with the length of the
observation period despite huge differences in the number of
earthquakes $N$ --- which is very different from a random process
\citep{davidsen06mp}. As Fig.~\ref{fig:l} shows, $P_{2.5}(l)$ is
largely unaltered if only the sub-catalog up to 1988 is analyzed.
This is not true for sub-catalogs of similar size generated by
randomly deleting events. The comparison of the two different
observation periods in Fig.~\ref{fig:l} further shows that $l^*$
does not depend strongly on the total number of recurrences (or
links) or on the average degree of the network, $\langle k
\rangle= {\rm \# links/\# nodes}$ ($\langle k \rangle = 6.56$
(7.40) for events up to 1988 (2002) and $m=2.5$), but clearly on
$m$. The independence of the time span and consequent number of
events implies that Eq.~(\ref{equ:p(l)}) is a robust, empirical
result for seismicity.

The identification $l^*=L_R$ is also consistent with the fact that
the description of earthquakes as a point process  breaks down  at
the rupture length. Below that scale, the relevant distance(s)
between earthquakes is not given solely by their epicenters but
also by the relative location and orientation of the spatially
extended ruptures. Due to different orientations we expect
randomness or lack of correlations between epicenters for
distances below the rupture length. If events are happening
randomly in space, or are recorded as happening randomly in space
due to location errors, then $P_m(l)$ rises linearly. To see this
consider a two dimensional disc of radius $R$, with one point at
the center and $N_R$ randomly distributed points. The probability
that there will be no (other) point within a distance $l$ of the
center point is $(1 -l^2/R^2)^{N_R}$; therefore, the probability
density for the closest point to be at distance $l$ is
$(2N_Rl/R^2)(1- l^2/R^2)^{N_R-1}$. At small $l$, this will
describe the distribution shown in Fig.~\ref{fig:l} and determine
the scaling function $F$ in Eq.~(\ref{equ:p(l)}). In fact, this is
precisely what the earthquake data show for distances smaller than
the rupture length (see the straight line with a slope of 2.05 in
the inset of Fig.~1 and the linear increase with slope 1 in the
main part of Fig.~\ref{fig:l}).

The lengths $l^*$ observed for the values of $m$ we consider are
larger than the length ($\approx 100 m$) at which we observe
random behavior due to location errors. In fact, the data do not
show any anomaly near $100 m$. Moreover, $P_4(l)$ (blue triangles)
does not change substantially if the epicenters in the catalog are
randomly relocated by a small distance up to one kilometer. Yet,
the maximum for $P_{2.5}(l)$ shifts to larger $l$ with this
procedure, destroying the scaling of $l^*(m)$. Since the smallest
$l^*$ that obeys the data collapse is $\approx 160$ m, the data
collapse we observe for the original data verifies that the
relative location errors are indeed less than $100 m$, or of that
order. Furthermore, our observations indicate that spatial
correlations between epicenters are already lost for distances
$100 {\rm m} < l < l^*$, although the frequency of pairs of
recurrent events with these small distances is much higher than by
random chance \citep{davidsen06mp}.
\footnote{Note that a systematic
dependence of the location error on magnitude has not been reported
in the literature and is also not present in the catalog at hand.
It is unlikely that the characteristic length we see ($l^*$) is merely
an artifact due to location error growing with magnitude.}

\begin{figure*}
\noindent\includegraphics*[width=39pc]{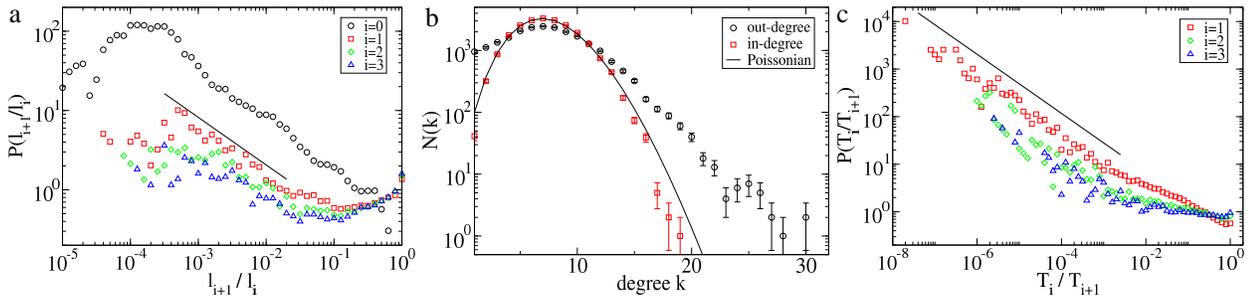}
\caption{\label{rank_ratio_l_all} (a) Distribution of recurrence
distance ratios $l_{i+1}/l_i$. The straight line corresponds to a
decay with exponent 0.6. \label{in_out_2} (b) Distributions of in-
and out-degrees of the network  for $m = 2.5$. The given error
bars correspond to $\sqrt{N(k)}$. \label{fig:ratio_t} (c)
Distribution of recurrence time ratios $T_{i}/T_{i+1}$. The
straight line has slope -0.62. }
\end{figure*}

Related to the distribution of distances of recurrent events is
the distribution of distance ratios $l_i/l_{i-1}$ in the cascade
of recurrences to a given event. Here recurrences are ordered by
time; recurrence $i$ comes after $i-1$. We take $l_0 = 448.5$ km,
which is the size of the region covered by the catalog
(Fig.~\ref{rank_ratio_l_all}a). By construction these ratios are
always $\leq 1$. We denote by $P_i(x)$ the probability density
that $l_i/l_{i-1}=x$ for each event that has an $i^{th}$
recurrence. The data for $i=1$ (black circles) scale over a wide
region as $P_1(x)\sim x^{-\delta_r}$ with $\delta_r \approx 0.6$
--  as already shown in \citep{davidsen05m}. This is indicated in
Fig.~\ref{rank_ratio_l_all}a by the straight line. Although each
distribution $P_i(x)$ is different, the curves for $i \geq 2$ also
show (more restricted) power law decay comparable to $P_1$. For
$l_{i+1}/l_i \to 1$ they also show a peak, which becomes more
pronounced with increasing $i$. This is due to recurrences
occurring at almost the same distance. The observed exponent
$\delta_r$ for the power law decay has a  dynamical origin and is
\emph{not} determined by the spatial distribution of seismicity
\citep{davidsen05m}: Purely based on the correlation dimension
$D_2$, one would expect $P_1(x)\sim x^{D_2-1}$. For Southern
California, this gives a growing dependence $P_1(x)\sim x^{0.2}$
rather than a decaying behavior. Thus, the exponent $\delta_r$
reflects the complex \emph{spatiotemporal} organization of
seismicity.

\begin{figure}
   \noindent\includegraphics*[width=20pc]{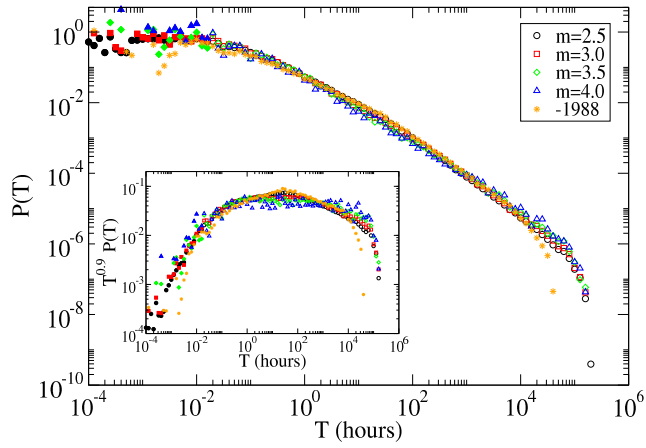}
   \caption{\label{fig:t} Distributions of recurrence times
       for different threshold magnitudes $m$. The distribution for $m=2.5$ up to 1988 is also shown.
       Filled symbols correspond
       to times below 90 seconds which are underestimated and
    unreliable due to measurement restrictions. The inset shows the rescaled distributions.}
   \end{figure}

A similar analysis can be made for the distribution of recurrence
times, $P_m(T)$ for different threshold magnitudes $m$, which is
shown in Fig.~\ref{fig:t}. These distributions all decay roughly
as $1/T^{\alpha}$ with $\alpha\approx 0.9$ for intermediate times
as indicated in the inset. The apparent scaling region in
Fig.~\ref{fig:t} shows some curvature, though. Surprisingly,
$P_m(T)$ is independent of $m$ and the number of events in the
considered catalog. This is very different from earlier results
for waiting time distributions between subsequent earthquakes
\citep{bak02,corral03} and reflects a new non-trivial feature of
the spatiotemporal dynamics of seismicity that appears when events
other than the immediately subsequent ones are considered.

The relative times between subsequent recurrences in the hierarchy
can be analyzed in the same way as distances were above.
Fig.~\ref{fig:ratio_t} shows distributions of ratios of the  times
$T_i/T_{i+1}$ for subsequent recurrences to a given event. The
broadest scaling regime materializes for $T_1/T_2$, again with
exponent $\delta_T \approx 0.6$. The distributions for larger $i$
follow roughly the same behavior for ratios $ \ll 1$, but deviate
(less strongly than for the spatial data in
Fig.~\ref{rank_ratio_l_all}a) when the ratios tend to 1. Again it
is obvious that this behavior cannot be explained by random
events.

The description of seismicity as a network of earthquake
recurrences allows its characterization by means of the usual
characteristics that are thought to be important for complex
networks \citep{albert02}. One such network property is its degree
distribution. Fig.~\ref{in_out_2}b shows the degree distributions
for $m = 2.5$, which is compared to a a Poisson distribution with
the same mean degree $\langle k \rangle = 7.40$ (solid line). A
Poisson degree distribution would be expected if earthquakes
epicenters were placed randomly in space and time. While the
in-degree distribution agrees with such a random network, the
out-degree distribution shows significant deviations. In
particular, the network keeps a preponderance of nodes with small
out-degree as well as an excess of nodes with large out-degree
compared to a Poisson distribution. This effect is independent of
magnitude, as an analysis of subsets with higher magnitude
threshold shows. Note, however, that $\langle k \rangle$ decreases
with $m$, simply because the catalog size shrinks with $m$. In
particular, we find $\langle k \rangle = 6.24, 5.20, 4.35$ for $m
= 3.0, 3.5, 4.0$, respectively. The non-trivial behavior of the
out-degree distribution implies in particular that the network
topology and, thus, the hierarchial cascade of recurrences or
records captures important information about the spatiotemporal
clustering of seismicity.
\footnote{Our results are robust with respect to modifications of
the rules used to construct the network, e.g., using spatial
neighborhoods such that the construction becomes symmetric under
time reversal or taking into account magnitudes. All such
modifications have the drawback that they do not define a record
breaking process consisting of recurrences to each event. Our
results are also unaltered if we exclude links with propagation
velocities larger than $6 km/sec$ ($\approx 0.1\%$ of all links).}

\section{Conclusions}

Our analysis shows that the description of seismicity by means of
recurrences in space-time allows us to characterize its clustering
behavior using only spatiotemporal relations between events and to
identify new, robust scaling laws in the pattern of seismic
activity. The pairs of recurrent events form a complex network
with non-trivial statistics. The method allows us to detect the
rupture length and its scaling with magnitude directly from
earthquake catalogs without making any assumptions. Our results
for the distributions of relative separations for the next
recurrence in space and time should also have implications for
seismic hazard assessment. Finally, our findings provide detailed,
benchmark tests for models of seismicity.

%
%

\begin{acknowledgments}
We thank the Southern California Earthquake Center (SCEC) for
providing the data.
\end{acknowledgments}

%
%

%
%

%
%

\end{article}

%
%
%
%
%
%

\end{document}